\documentclass[conference]{IEEEtran}
\IEEEoverridecommandlockouts

\usepackage{cite} 
\usepackage{amsmath,amssymb,amsfonts}
\usepackage{algorithmic}
\usepackage{graphicx} 
\usepackage{textcomp}
\usepackage{xcolor}
\usepackage[caption=false,font=footnotesize]{subfig}
\usepackage{bm} 
\usepackage[bookmarks=false, draft]{hyperref} 
\usepackage{multirow} 
\usepackage{booktabs} 
\usepackage{placeins} 
\usepackage{balance} 
\usepackage{footnote} 
\usepackage{tabularx} 
\usepackage{stfloats} 
\usepackage{cleveref}

\begin{document}

\title{SemCSINet: A Semantic-Aware CSI Feedback Network in Massive MIMO Systems}

\author{Ruonan Ren, Jianhua Mo and Meixia Tao\\
Department of Electronic Engineering,
Shanghai Jiao Tong University, Shanghai, China\\
\{ronan4332, mjh, mxtao\}@sjtu.edu.cn}

\maketitle

\begin{abstract}
Massive multiple-input multiple-output (MIMO) technology is a key enabler of modern wireless communication systems, which demand accurate downlink channel state information (CSI) for optimal performance. Although deep learning (DL) has shown great potential in improving CSI feedback, most existing approaches fail to exploit the semantic relationship between CSI and other related channel metrics.
In this paper, we propose SemCSINet, a semantic-aware Transformer-based framework that incorporates Channel Quality Indicator (CQI) into the CSI feedback process. By embedding CQI information and leveraging a joint coding-modulation (JCM) scheme, SemCSINet enables efficient, digital-friendly CSI feedback under noisy feedback channels.
Experimental results on DeepMIMO datasets show that SemCSINet significantly outperforms conventional methods, particularly in scenarios with low signal-to-noise ratio (SNR) and low compression ratios (CRs), highlighting the effectiveness of semantic embedding in enhancing CSI reconstruction accuracy and system robustness.

\end{abstract}

\begin{IEEEkeywords}
CSI feedback, deep learning, massive MIMO, channel quality indicator (CQI)
\end{IEEEkeywords}

\section{Introduction}

\begin{figure*}[h]
\centering
\includegraphics[width=7in]{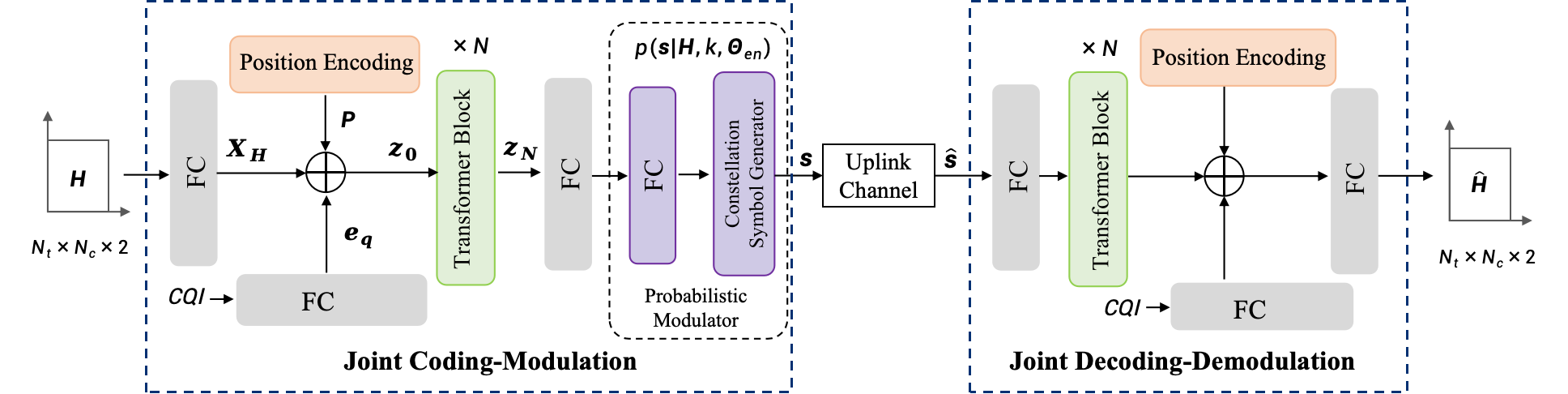}
\caption{Architecture of the proposed SemCSINet.}
\label{fig:framework}
\end{figure*}

Massive multiple-input multiple-output (MIMO) technology has emerged as a cornerstone of 5G networks, supporting key service categories such as enhanced Mobile Broadband (eMBB). It is also expected to play a pivotal role in shaping future 6G systems, such as the immersive communication, and integrated sensing and communication \cite{mimo_6g}. To fully exploit the potential of MIMO, highly accurate downlink channel state information (CSI) is indispensable \cite{mimo_benefit}. 


As 6G systems continue to scale up the number of antennas, the dimensionality of CSI grows proportionally, thereby imposing significant transmission overhead. This challenge is particularly pronounced in frequency division duplex (FDD) systems, where downlink CSI must be explicitly fed back to the base station, leading to increased complexity and uplink resource consumption.

To alleviate this burden, prior work has explored three principal categories of CSI feedback methods: codebook-based methods \cite{csi_codebook}, compressive sensing (CS)-based methods \cite{csi_cs}, and deep learning (DL)-based methods \cite{guo_ai_2024, Guo_Yiran_COMM24}. 
Among these, DL-based methods have demonstrated superior compression efficiency with minimal loss in reconstruction accuracy. Notably, CsiNet \cite{csinet} and CsiNet+ \cite{csinet+} established the viability of neural networks for CSI feedback. Later advancements, including attention mechanisms (e.g., CsiNet-LSTM \cite{LSTM_csinet}, Attention-CSI \cite{attention_csinet}, TransNet \cite{transnet}) and Transformer-based architectures (e.g., SwinCFNet \cite{swincfnet}), have further improved performance. In parallel, lightweight models such as CRNet \cite{crnet} and CLNet \cite{clnet} have reduced computational complexity, enhancing the practical deployment of DL-based CSI feedback solutions.

Despite these advances, most existing methods treat CSI values in isolation, neglecting their semantic correlations with auxiliary channel metrics such as the Channel Quality Indicator (CQI). CQI, which is a metric derived from channel gain at the user side, is commonly fed back to the base station at low cost \cite{3gpp_cqi}, and encapsulates rich semantic information about the underlying channel quality. However, integrating CQI into CSI feedback has remained largely unexplored. Moreover, many DL-based methods adopt analog transmission, wherein the neural network output is directly mapped to channel inputs \cite{jscc}, making them ill-suited for practical digital communication systems.

In this paper, we propose \textbf{SemCSINet}, a semantic-aware deep learning framework designed to enhance CSI feedback for massive MIMO systems. Distinct from prior approaches, SemCSINet explicitly incorporates CQI information into the CSI encoding process, leveraging the strong semantic link between CSI and CQI to improve reconstruction accuracy. In addition, we integrate a joint coding-modulation (JCM) framework \cite{jcm}, enabling the quantized and modulated feedback to remain robust under noisy channel conditions while maintaining compatibility with digital transmission systems.

Experimental results demonstrate that SemCSINet substantially outperforms conventional feedback methods, particularly under challenging scenarios characterized by low signal-to-noise ratios (SNRs) and low compression ratios (CRs). These findings highlight the effectiveness of semantic embedding for CSI reconstruction and offer a viable path toward practical deployment in future wireless networks.

\section{System Model}

\subsection{CSI Feedback}

We consider a single-cell FDD massive MIMO system, where the base station (BS) with $N_t$ transmitting antennas serves the single-antenna user equipment (UE) over $N_c$ subcarriers.
The CSI matrix in the spatial-frequency domain can be expressed in the matrix form as:
\begin{equation}
\mathbf{H} = [\mathbf{h}_1, \mathbf{h}_2, \dots, \mathbf{h}_{N_c}] \in \mathbb{C}^{N_t \times N_c}.
\end{equation}

The BS requires estimates of the downlink CSI matrix to design the precoding vector. In the FDD system, the UE first estimates the downlink channel. Then, the estimated CSI matrix $\mathbf{H}$ is compressed and sent to the BS in the uplink channel. We assume that the UE can perfectly estimate $\mathbf{H}$ in this paper.

\subsection{CQI Feedback}

In massive MIMO orthogonal frequency-division multiplexing (OFDM) systems, the CQI feedback plays a crucial role in determining the modulation and coding scheme (MCS) of the downlink transmission. To reduce feedback overhead, two approaches are commonly employed: wideband CQI and subband CQI. Both approaches aim to efficiently represent the channel conditions and optimize the transmission performance.

\subsubsection{Wideband CQI Feedback}

In the wideband CQI feedback approach, the CQI is determined based on the average SNR over all subcarriers, taking into account the overall channel quality rather than individual subcarriers.

The wideband SNR, denoted as $\rho_{w}$, is computed as the average of the SNRs over all the $N_c$ subcarriers:
\begin{equation} 
\rho_{w} = \frac{1}{N_c}\sum_{n=1}^{N_c} \rho_n, 
\end{equation} 
where $\rho_n$ represents the SNR of the $n$-th subcarrier.

The CQI index $\mathbf{k}_{w}$, ranging from 0 to 15, is then selected by looking up the corresponding value in the CQI-SNR mapping table \cite{cqi_mapping}, which associates a specific SNR value with the corresponding modulation and coding scheme. 

\subsubsection{Subband CQI Feedback}

In contrast, subband CQI feedback reports the channel quality for each subcarrier group separately, enabling finer-grained link adaptation. By computing CQI based on subband-specific SNR, it more accurately captures frequency-selective fading, allowing the system to optimize transmission per subband instead of relying on a coarse wideband estimate. This leads to improved performance in channels with significant subband variability.

For each subband $m$, the SNR $\rho_m$ is computed as the average SNR over the subcarriers belonging to the subband:

\begin{equation} 
\rho_m = \frac{1}{N_m} \sum_{n \in \mathcal{N}_m} \rho_n, 
\end{equation} where $\mathcal{N}_m$ represents the set of subcarriers belonging to subband $m$, and $N_m$ is the number of subcarriers in subband $m$.

Each subband CQI index $\mathbf{k}_s$ is then determined by referring to the corresponding SNR value for that subband in the CQI-SNR mapping table \cite{cqi_mapping}. Assuming there are $N_b$ subbands, the subband CQI is then a vector $\mathbf{k}_s = [k_{s,1}, k_{s,2}, \dots, k_{s,N_b}]$.

\subsection{DL-based CQI-Assisted CSI Feedback}

In this section, we elaborate on the CQI-assisted CSI feedback process over the uplink channel. Since CQI encapsulates high-level semantics about channel conditions, it is leveraged as auxiliary information to improve the fidelity and robustness of compressed CSI feedback. Specifically, the CQI index is embedded into the encoding and modulation stages, allowing the system to adaptively enhance the semantic representation of the CSI matrix $\mathbf{H}$ under varying channel environments.

At the UE, the channel matrix $\mathbf{H}$ and the CQI value $\mathbf{k}$ ($\mathbf{k}_w$ for wideband or $\mathbf{k}_s$ for subband) are encoded and modulated using a JCM process. 

The output of the joint-encoder-modulator is a constellation sequence $\mathbf{s} = [s_1, s_2, \dots, s_M] \in \mathbb{C}^M$, where $M$ represents the number of channel uses. Each element in the sequence $\mathbf{s}$ takes values from a predefined constellation map $\mathcal{C}$.
The encoding and modulation process can be expressed as: 
\begin{equation} 
\mathbf{s} = f(\mathbf{H}, \mathbf{k}, \Theta_{\text{en}}), 
\end{equation} 
where $\Theta_{\text{en}}$ denotes the network parameters at the UE. The compression rate is defined as $\gamma = \frac{M}{N_t N_c}$.

Once the constellation sequence $\mathbf{s}$ is generated, it is transmitted over the uplink additive white Gaussian noise (AWGN) channel, which can be expressed as: 
\begin{equation} 
\hat{\mathbf{s}} = h(\mathbf{s}) = \mathbf{s} + \boldsymbol{\epsilon}, 
\end{equation} 
where $\boldsymbol{\epsilon} \sim \mathcal{CN}(\mathbf{0}, \sigma^2 \mathbf{I}_{M})$ is the channel noise, 
with each element independently drawn from a complex Gaussian distribution with zero mean and variance $\sigma^2$. The uplink feedback channel condition is characterized by the channel SNR.

At the BS, the received constellation sequence $\hat{\mathbf{s}}$ is demodulated and decoded to recover the original CSI matrix $\mathbf{H}$. The decoding and demodulation process is given by: 
\begin{equation} 
\widehat{\mathbf{H}} = g(\hat{\mathbf{s}}, \mathbf{k}, \Theta_{\text{de}}), \end{equation} 
where $\Theta_{\text{de}}$ represents the network parameters at the BS, and $\widehat{\mathbf{H}} \in \mathbb{C}^{N_t \times N_c}$ denotes the recovered CSI matrix.

The end-to-end training approach is employed to optimize the SemCSINet architecture. The loss function is defined as the mean squared error (MSE) between the reconstructed CSI matrix and its ground truth: 
\begin{align} L(\Theta_{\text{en}}, \Theta_{\text{de}}) &= \frac{1}{T} \sum_{i=1}^{T} \| \widehat{\mathbf{H}}_i - \mathbf{H}_i \|_F^2 \\
&=\frac{1}{T} \sum_{i=1}^{T} \| g\left(h\left(f(\mathbf{H}_i, \mathbf{k}_i, \Theta_{\text{en}}), \mathbf{k}_i, \Theta_{\text{de}}\right)\right)-\mathbf{H}_i \|_F^2, \end{align} 
where $T$ is the total number of samples in the training set, $i$ denotes the $i$-th sample, and $\| \cdot \|_F$ is the Frobenius norm.

\section{Design of SemCSINet}
\label{sec:design}

The overall architecture of SemCSINet is illustrated in Fig. \ref{fig:framework}, which enables efficient integration of CSI and CQI features, with the learned positional encoding helping to preserve essential information regarding subcarrier and antenna positions.

SemCSINet processes the input channel matrix $\mathbf{H} \in \mathbb{C}^{N_t \times N_c}$ through a fully connected (FC) layer to generate the channel feature vectors $\mathbf{X}_H \in \mathbb{R}^{N_{\text{embed}}}$. In parallel, the CQI index $\mathbf{k}$ is embedded through another FC layer, producing CQI feature vectors $\mathbf{e}_q \in \mathbb{R}^{N_{\text{embed}}}$. These two feature vectors, $\mathbf{X}_H$ and $\mathbf{e}_q$, are then combined element-wise to integrate the channel and CQI information.

To preserve the spatial information across subcarriers and antennas, the combined feature vector is augmented with a position encoding. In our implementation, the position encoding is treated as a learnable parameter, denoted as $\mathbf{P} \in \mathbb{R}^{N_{\text{embed}}}$, which is added to the combined feature vector. The resulting feature vector, with positional information, is expressed as:
\begin{equation}
\mathbf{z}_0 = \mathbf{X}_H + \mathbf{e}_q + \mathbf{P}.
\end{equation}
The position-encoded feature vector $\mathbf{z}_0$ is then passed through a series of $N$ transformer blocks \cite{transformer}. The output of the transformer after $N$ layers is denoted as $\mathbf{z}_N \in \mathbb{R}^{N_{\text{embed}}}$.
Then, the probabilistic modulator \cite{jcm} generates the transition probability distribution $p_{\text{en}}(\mathbf{s}|\mathbf{H}, \mathbf{k}, \Theta_{\text{en}})$. 
The constellation symbol generator produces the final discrete modulation symbols $\mathbf{s}$, which are transmitted over the uplink channel.

At the BS side, the decoding and demodulation process mirrors the encoding and modulation procedure, involving estimating the posterior distribution $p_{\text{de}}(\mathbf{H}|\mathbf{\hat{s}}, \mathbf{k}, \Theta_{\text{de}})$. 
After passing  $\mathbf{\hat{s}}$ through a FC layer, then $N$ transformer blocks, position encoding module, and another FC layer, the BS reconstructs the estimated CSI, denoted as $\widehat{\mathbf{H}}$. 

\section{Experimental Results}
\label{sec:experiments}

In this section, we evaluate the performance of the proposed SemCSINet framework through comprehensive experiments conducted on DeepMIMO dataset~\cite{deepmimo1}, \cite{deepmimo2}. 

We compare the performance of our proposed JCM-based method (denoted as JCM) under three different CQI embedding configurations: with wideband CQI (denoted as wide\_CQI), with subband CQI (denoted as sub\_CQI), and without CQI embedding (denoted as wo\_CQI). Additionally, experiments are conducted without the JCM framework to verify the performance of CQI embedding under different transmission schemes. Specifically, the output of the semantic encoder is directly sent into the channel without probabilistic modulation. Therefore, this method is called the analog modulation method (denoted as Analog).

These experiments are conducted under various feedback channel SNR conditions and compression ratios, demonstrating the advantages of CQI embedding in improving system performance.

\subsection{Experimental Setup} \label{subsec:setup}
\begin{table}[t]
\renewcommand{\arraystretch}{1.3}  
\caption{System Configuration for Datasets}
\label{tab:parameters}
\resizebox{0.46\textwidth}{!}{  
\begin{tabular}{|c|c|}
\hline
\textbf{Parameter} & \textbf{Value} \\ \hline 
Frequency & 3.5 GHz  \\ \hline
Subcarrier Spacing & 30 kHz  \\ \hline
Height of BS & 6 m \\ \hline
Height of UEs & 2 m \\ \hline
BS Transmit Power & 30 dBm  \\ \hline
BS Antenna Configuration & 4V8H1P array \\ \hline
UE Antenna Configuration & single isotropic \\ \hline
Effective Antenna Elements ($N_t$) & 32 \\ \hline
Number of RBs & 52  \\ \hline
Sampled Subcarriers ($N_c$) & 52 \\ \hline
Number of Sampled Subcarriers per Subband & 4  \\ \hline
\end{tabular}
}
\end{table}

\begin{figure}[t]
\centering
\subfloat[Top view of the Los Angeles scenario.]{%
  \includegraphics[width=0.3\textwidth]{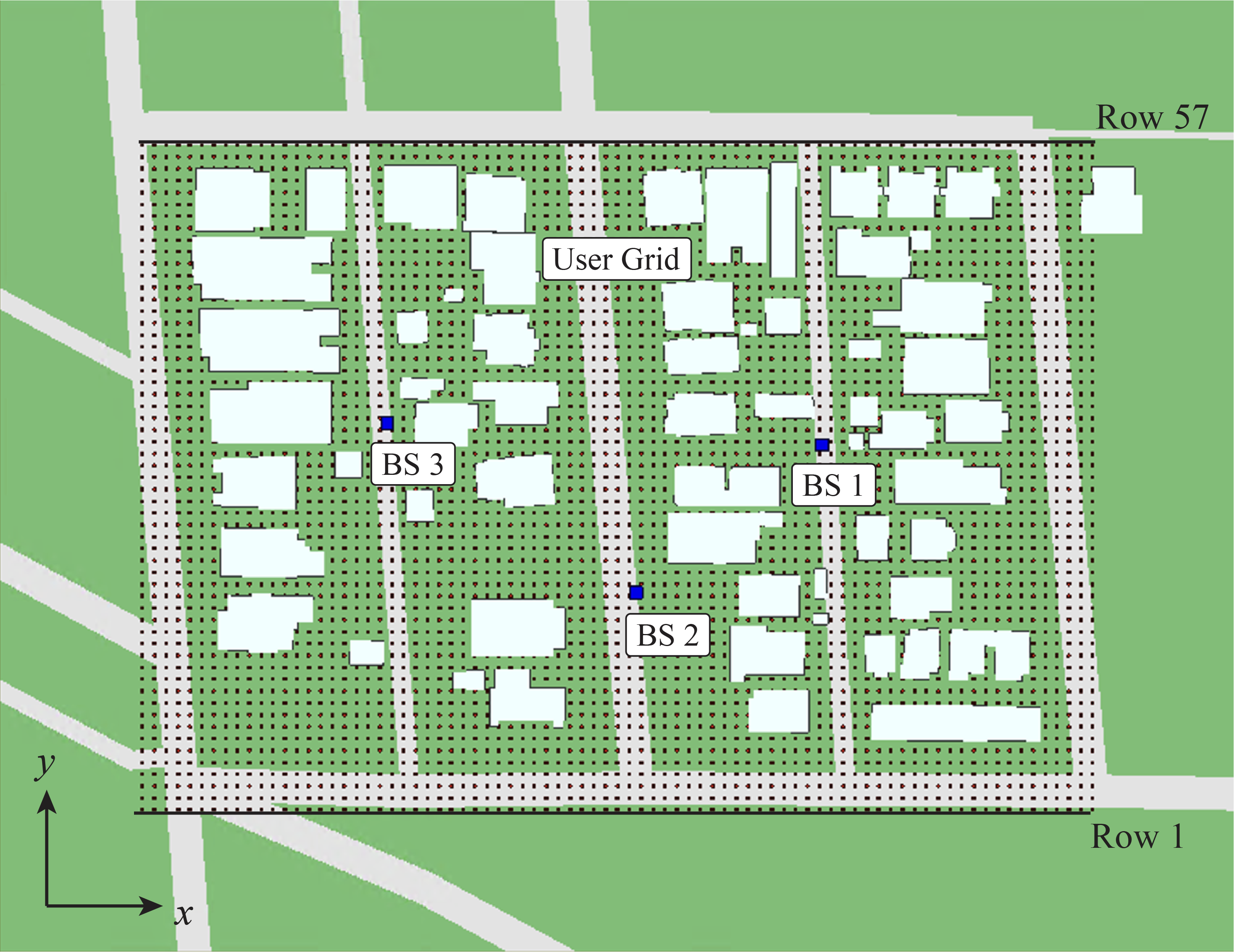}%
  \label{fig:LA_topview}}
\hfil
\subfloat[Wideband CQI distribution of the Los Angeles scenario.]{%
  \includegraphics[width=0.45\textwidth]{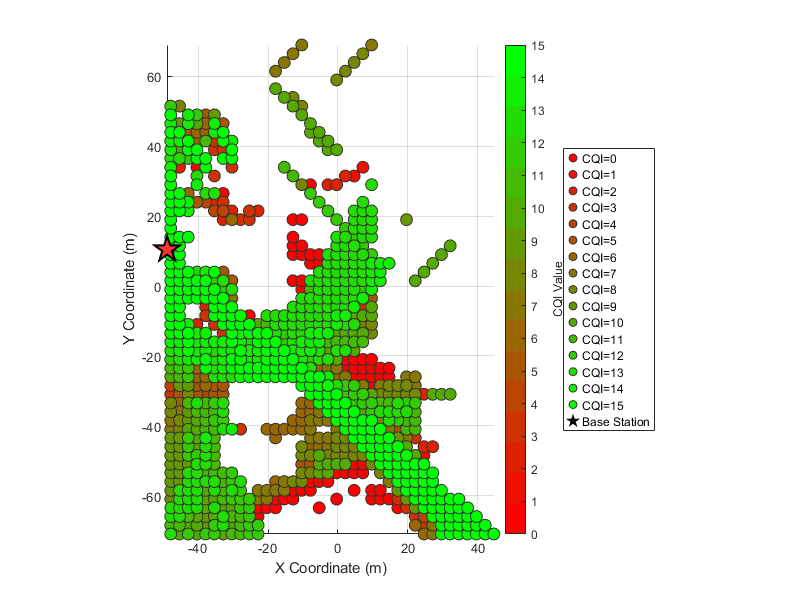}%
  \label{fig:LA_CQI_distribution}}
\caption{Simulation of the Los Angeles scenario in the DeepMIMO dataset.}
\label{fig:LA_sim}
\end{figure}

\begin{figure}[t]
\centering
\subfloat[Top view of the San Francisco scenario.]{%
  \includegraphics[width=0.28\textwidth]{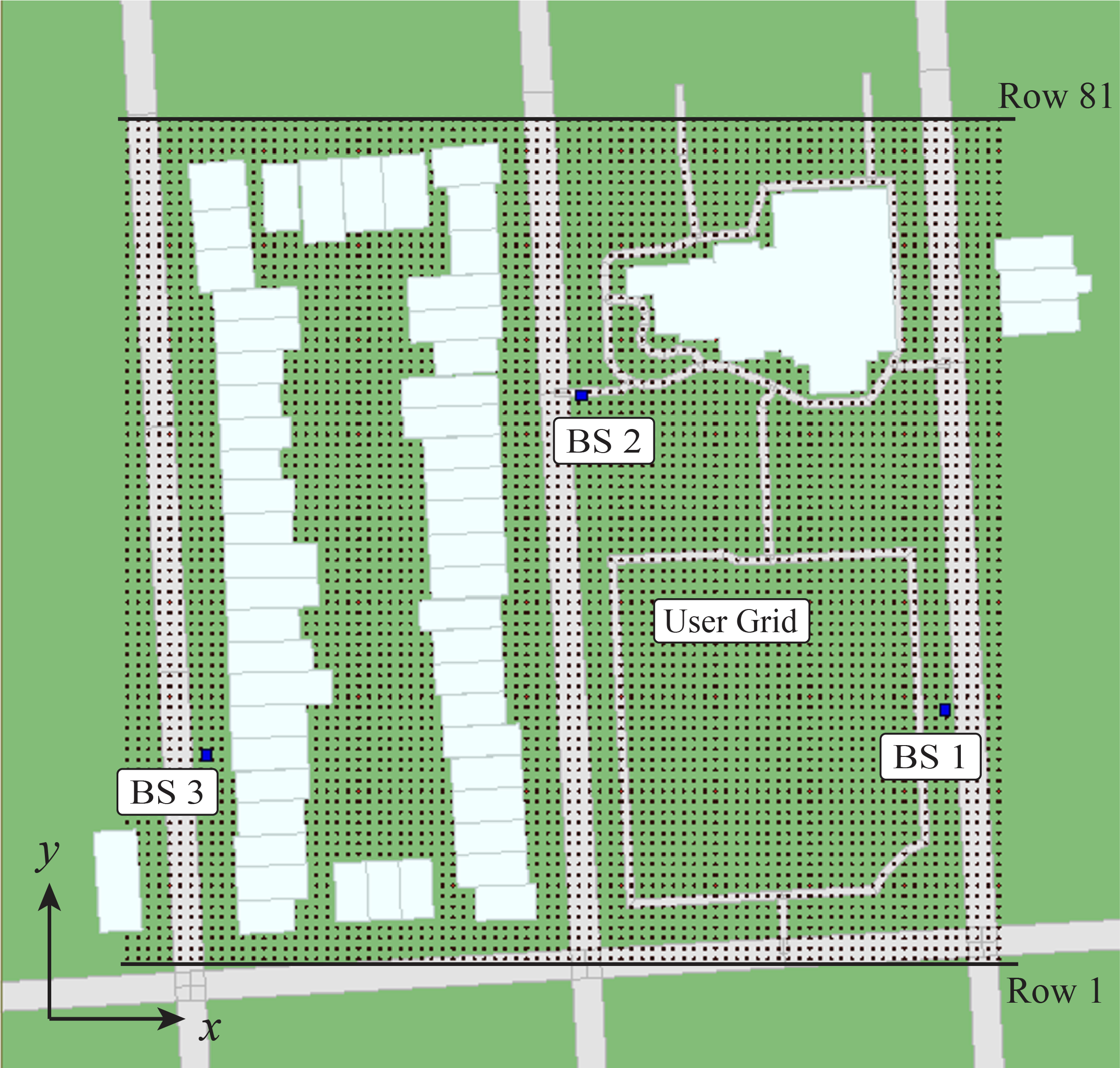}%
  \label{fig:SF_topview}}
\hfil
\subfloat[Wideband CQI distribution in the San Francisco scenario.]{%
  \includegraphics[width=0.32\textwidth]{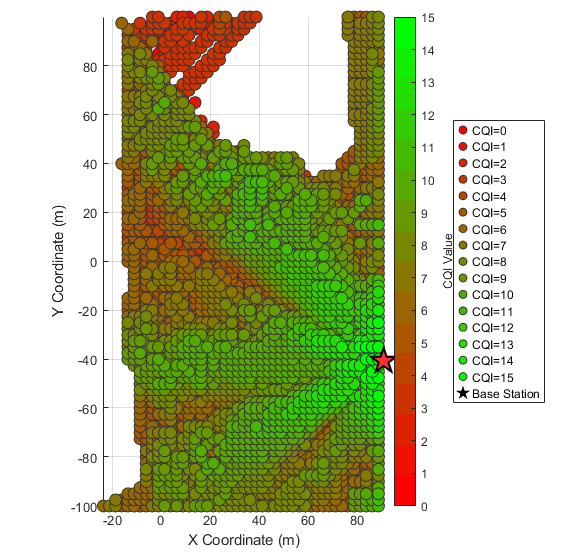}%
  \label{fig:SF_CQI_distribution}}
\caption{Simulation of the San Francisco scenario in the DeepMIMO dataset.}
\label{fig:SF_sim}
\end{figure}

We evaluate the impact of CQI embedding using the DeepMIMO dataset under realistic urban scenarios. Specifically, we sample the first subcarrier of each resource block (RB), resulting in $N_c = 52$ representative subcarriers. Key simulation parameters are summarized in Table~\ref{tab:parameters}. The BS is equipped with $N_t = 32$ single-polarized antennas, arranged as 4 vertical and 8 horizontal elements.

Two scenarios from the DeepMIMO dataset are considered: Los Angeles and San Francisco. Fig.~\ref{fig:LA_topview} shows the top view of the Los Angeles scenario, where BS 3 is activated and users located to its east are selected. The corresponding wideband CQI distribution is shown in Fig.~\ref{fig:LA_CQI_distribution}. For the San Francisco scenario, Fig.~\ref{fig:SF_topview} presents the top view, with BS 1 activated and users to its west selected. The wideband CQI distribution for this case is shown in Fig.~\ref{fig:SF_CQI_distribution}. Note that UEs for which no propagation paths were found in ray tracing~\cite{deepmimo2} are excluded from the CQI visualizations.



\subsection{Performance Comparison}

We evaluate the performance of SemCSINet using two metrics: normalized mean squared error (NMSE), and squared generalized cosine similarity (SGCS). These metrics are defined as follows:

\begin{equation}
    \text{NMSE} = \mathbb{E} \left\{\frac{\| \mathbf{H} - \hat{\mathbf{H}} \|_F^2}{\| \mathbf{H} \|_F^2} \right\},
    \label{eq:nmse}
    \end{equation}

    \begin{equation}
    \text{SGCS} = \frac{1}{N_c} \sum_{n=1}^{N_c} \left( \frac{|\hat{\mathbf{h}}_n^H \mathbf{h}_n|}{\|\hat{\mathbf{h}}_n\| \|\mathbf{h}_n\|} \right)^2,
    \label{eq:sgcs}
    \end{equation}
where $\hat{\mathbf{h}}_n$ and $\mathbf{h}_n$ are the recovered and true CSI vectors for the $n$-th subcarrier, respectively. 


\begin{figure}[t]
\centering
\includegraphics[width=0.42\textwidth]{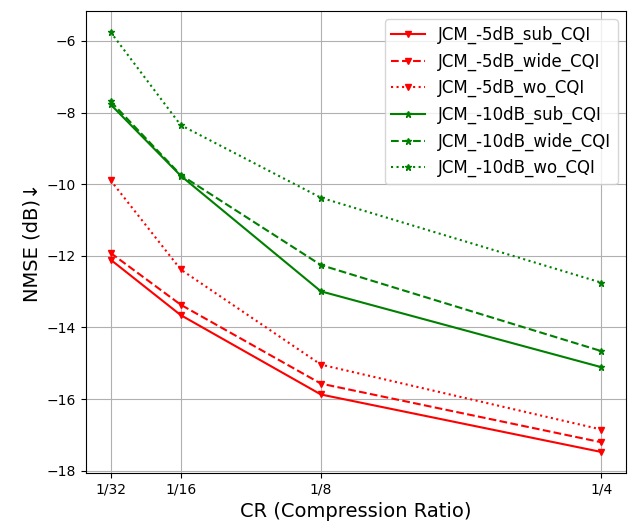}
\caption{NMSE performance under different CQI embedding configurations in the Los Angeles scenario dataset.}
\label{fig:LA-nmse}
\end{figure}

\begin{figure}[t]
\centering
\includegraphics[width=0.42\textwidth]{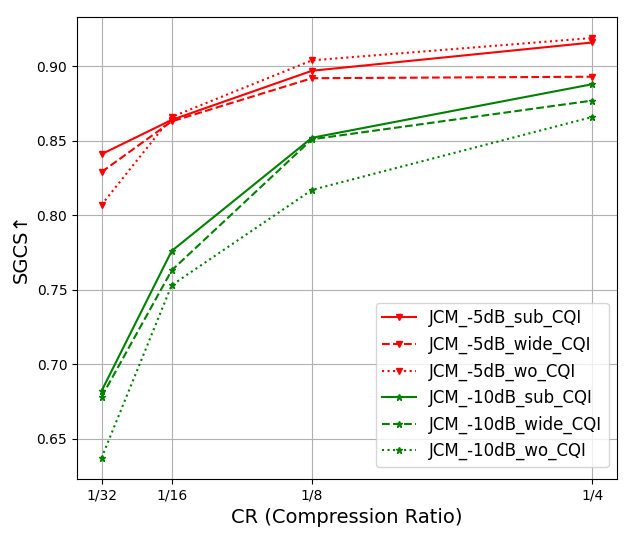}
\caption{SGCS performance under different CQI embedding configurations in the Los Angeles scenario dataset.}
\label{fig:LA-sgcs}
\end{figure}

\subsubsection{Performance on the Los Angeles Scenario Dataset}

Fig. ~\ref{fig:LA-nmse} and Fig. \ref{fig:LA-sgcs} present the simulation results for the Los Angeles scenario dataset, evaluating performance at uplink SNRs of $-10$~dB and $-5$~dB under various compression ratios.
The results clearly demonstrate that incorporating CQI embedding improves system performance in terms of both NMSE and SGCS metrics, especially under low SNR and low compression ratio conditions. At an SNR of $-10$~dB and a CR of $1/16$, the inclusion of subband CQI achieves a $2.6$~dB reduction in NMSE and a $4.3\%$ improvement in SGCS compared to the baseline model without CQI. In comparison, wideband CQI offers a more modest enhancement, with a $1.9$~dB reduction in NMSE and a $4.2\%$ gain in SGCS. These results highlight the critical role of CQI embedding in enhancing CSI feedback accuracy, with subband CQI proving more effective than wideband CQI.

\begin{figure}[t]
\centering
\includegraphics[width=0.41\textwidth]{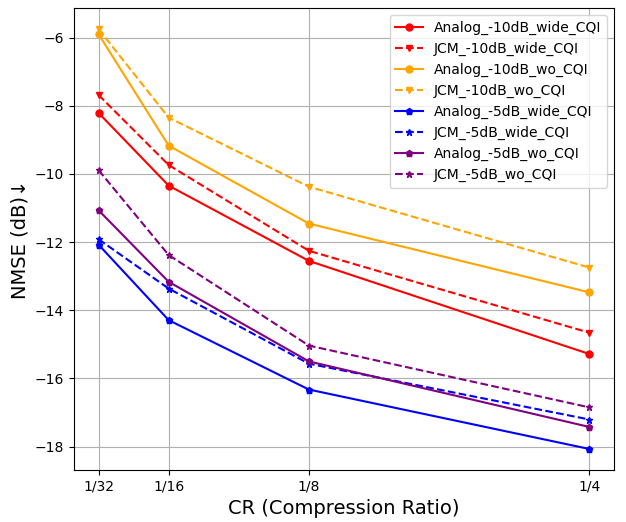}
\caption{NMSE performance under different transmission schemes in the Los Angeles scenario dataset.}
\label{fig:LA-nmse-ana}
\end{figure}


Fig.~\ref{fig:LA-nmse-ana} further illustrates the NMSE performance of different transmission schemes in the Los Angeles scenario. Across all schemes, incorporating wideband CQI consistently improves CSI feedback accuracy. Notably, the JCM-based method achieves performance close to the upper bound achieved by Analog, while enabling digital-friendly transmission. This makes it particularly suitable for practical deployment, as it strikes an effective balance between reconstruction accuracy, robustness to noise, and compatibility with existing digital communication infrastructure.

\subsubsection{Performance on the San Francisco Scenario Dataset}

\begin{figure}[t]
\centering
\includegraphics[width=0.42\textwidth]{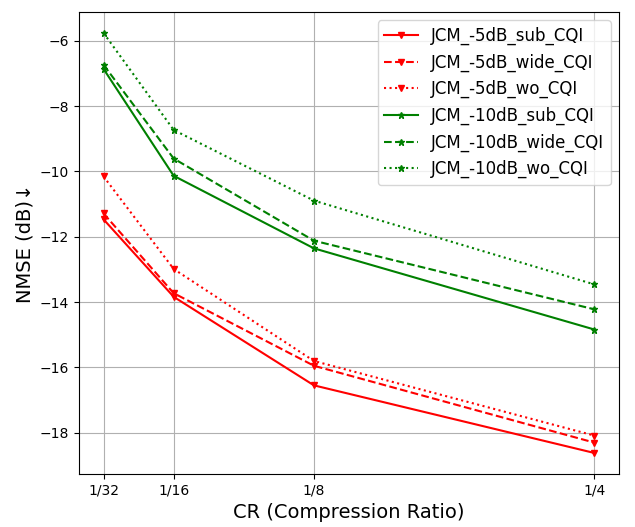}
\caption{NMSE performance under different CQI embedding configurations in the San Francisco scenario dataset.}
\label{fig:SF-nmse}
\end{figure}

Fig. \ref{fig:SF-nmse} shows the NMSE performance in the San Francisco scenario. 
The models with CQI embedding consistently achieve lower NMSE compared to those without CQI across all CR settings.
The most significant improvement is observed at a CR of $1/16$ and an SNR of $-10$~dB, where the subband CQI embedding model achieves a reduction of up to $1.5$~dB in NMSE.
However, the overall performance gain from CQI embedding in this scenario is less significant compared to the Los Angeles scenario, indicating that the effectiveness of CQI embedding may vary depending on the environmental characteristics of each scenario.




\subsubsection{Mechanism Analysis}
\begin{table}[t]
\renewcommand{\arraystretch}{1.3} 
\caption{Mutual Information and Entropy for CQI Embedding}
\label{tab:entropy}
\centering
\begin{tabular}{|c|c|c|c|c|}
\hline
\textbf{Metric} & \multicolumn{2}{c|}{\textbf{Wideband CQI}} & \multicolumn{2}{c|}{\textbf{Subband CQI}} \\ \hline
Scenario & \textbf{LA} & \textbf{SF} & \textbf{LA} & \textbf{SF} \\ \hline
Entropy of CQI (bits) & 3.43 & 3.31 & 3.77 & 5.46 \\ \hline
Mutual Information $I(\mathbf{H'}; \mathbf{k})$ (bits) & 3.01 & 2.66 & 3.06 & 2.81 \\ \hline
\end{tabular}
\end{table}

To further evaluate the effectiveness of CQI embedding, we provide a theoretical analysis based on entropy and mutual information. The results are summarized in Table~\ref{tab:entropy}.

In this context, higher entropy indicates that the feedback captures more detailed information, thereby better characterizing channel variations. We estimate the entropy of CQI using the k-nearest neighbors (k-NN) method \cite{kNN}, which is well-suited for continuous-valued data. As shown in Table~\ref{tab:entropy}, subband CQI consistently exhibits higher entropy than wideband CQI across both scenarios. Notably, in the San Francisco scenario, subband CQI reaches an entropy of 5.46 bits, which is 2.15 bits higher than that of wideband CQI, indicating a richer representation of frequency-selective channel features.

We further compute the mutual information between the CQI vector $\mathbf{k}$ and the normalized channel matrix $\mathbf{H}' = \frac{\mathbf{H}}{\|\mathbf{H}\|_F}$, again using the k-NN method. Normalization removes amplitude information, consistent with the nature of CSI feedback. Results show that subband CQI achieves slightly higher mutual information than wideband CQI in both scenarios. For instance, in the San Francisco case, subband CQI achieves 2.81 bits, compared to 2.66 bits for wideband CQI. These results confirm that finer-grained CQI embedding preserves more relevant channel information, facilitating more accurate CSI reconstruction at the BS.

\begin{figure}[t]
\centering
\includegraphics[width=0.5\textwidth]{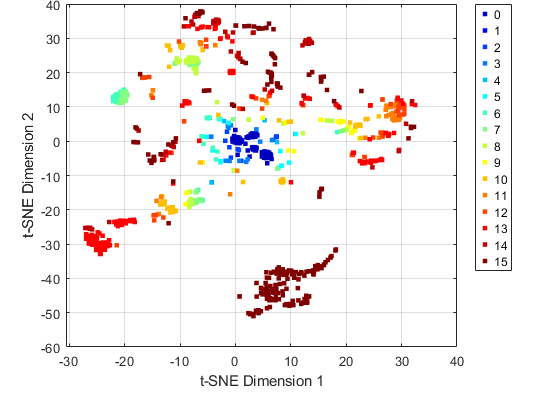}
\caption{t-SNE visualization of the CSI matrix in the Los Angeles scenario, with user samples color-coded by wideband CQI levels.}
\label{fig:LA-tsne}
\end{figure}

\begin{figure}[t]
\centering
\includegraphics[width=0.45\textwidth]{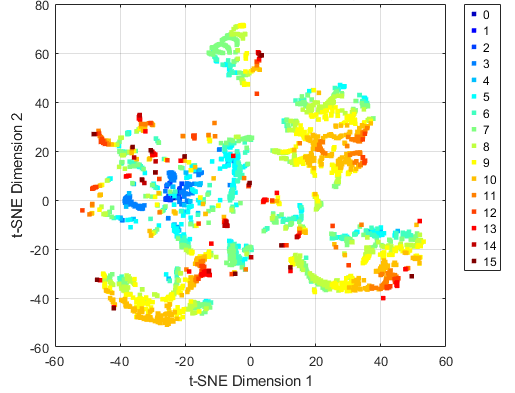}
\caption{t-SNE visualization of the CSI matrix in the San Francisco scenario, with user samples color-coded by wideband CQI levels.}
\label{fig:SF-tsne}
\end{figure}

We further assess the effectiveness of CQI embedding through t-distributed Stochastic Neighbor Embedding (t-SNE) \cite{t-SNE}, which visualizes the structure of high-dimensional CSI representations across different wideband CQI levels, as shown in Fig. \ref{fig:LA-tsne} and Fig. \ref{fig:SF-tsne}.
In the Los Angeles scenario, the CSI embeddings form well-separated clusters, indicating a strong correlation between CQI values and underlying channel states, consistent with the observed performance gains. In contrast, the San Francisco scenario exhibits greater cluster overlap, suggesting weaker CQI-channel correlation and limited benefit from CQI embedding.
These observations imply that the effectiveness of CQI embedding depends on the degree of channel state separability induced by CQI levels. This highlights the importance of scenario-specific CQI-channel relationships when designing feedback strategies.

\section{Conclusion}
\label{sec:conclusion}
In this paper, we have shown that incorporating CQI information, particularly subband CQI, substantially enhances CSI feedback performance in massive MIMO systems, especially under low SNR and low compression ratio conditions. In addition to empirical results, theoretical analysis based on entropy and mutual information, together with t-SNE visualizations, further validates the effectiveness of semantic embedding. These findings provide valuable insights and lay a solid foundation for advancing semantic-aware CSI feedback in practical wireless communication systems.


\bibliographystyle{IEEEtran} 
\bibliography{reference} 

\end{document}